\documentclass[onecolumn]{article}
\pdfoutput=1

\usepackage{jheppub,bm, amsmath,amssymb,amsfonts,slashed,array,graphicx}

\newcommand{\Gc}{G_{\textrm{c}}}
\newcommand{\GF}{G_{\textrm{F}}}
\newcommand{\GSM}{G_{\textrm{SM}}}
\newcommand{\GFp}{G_{\textrm{F}}^\prime}
\newcommand{\UF}{\textrm{U}(1)_{\textrm{F}}}

\newcommand{\UY}{\textrm{U}(1)_Y}

\newcommand{\Mpl}{M_{\rm pl}}
\newcommand{\gs}{g_{*_{\rm S}}}

\usepackage{xcolor}
\definecolor{red}{rgb}{1.0, 0, 0}

\begin{document}
\title{KeV Warm Dark Matter and Composite Neutrinos}
\author{Dean J. Robinson}
\author{Yuhsin Tsai}
\affiliation{Laboratory for Elementary-Particle Physics, Cornell University, Ithaca, N.Y.}
\emailAdd{djr233@cornell.edu}
\emailAdd{yt237@cornell.edu}
\date{\today}

\abstract{
Elementary keV sterile Dirac neutrinos can be a natural ingredient of the composite neutrino scenario. For a certain class of composite neutrino theories, these sterile neutrinos naturally have the appropriate mixing angles to be resonantly produced warm dark matter (WDM). Alternatively, we show these sterile neutrinos can be WDM produced by an entropy-diluted thermal freeze-out, with the necessary entropy production arising not from an out-of-equilibrium decay, but rather from the confinement of the composite neutrino sector, provided there is sufficient supercooling.}

\maketitle

\section{Introduction} 
Sterile neutrinos with masses at the keV scale are a popular warm dark matter (WDM) candidate \cite{Olive:1982cb,Dodelson:1994sn,Shi:1998km,Fuller:2001sn,Dolgov:2000ew,Biermann:2006bu,Boyarsky:2006fg,Boyanovsky:2006it,Asaka:2007ls, Sierra:2008wj,Laine:2008pg,Wu:2009yr,Gelmini:2009xd,Boyarsky:2009ix,Kusenko:2009up,deVega:2009ku,deVega:2010yk,deVega:2011si, Araki:2011zg,Chen:2011ai,Merle:2011yv,Geng:2012jm}, that may potentially account for small-scale structure formation (see e.g \cite{Bode:2000gq,Zavala:2009ms, deVega:2010yk}) and possibly explain large pulsar kick velocities \cite{Kusenko:1997sp,Kusenko:2009up}. Sterile neutrino WDM can be produced non-thermally via (non)-resonant oscillations from the active neutrinos \cite{Dodelson:1994sn,Shi:1998km,Fuller:2001sn,Boyarsky:2007ay,Boyarsky:2008xj,Wu:2009yr,Kusenko:2009up, Boyarsky:2009ix,Das:2010ts,Watson:2011dw}, by decays from the inflaton \cite{Shaposhnikov:2006xi,Anisimov:2008qs}, or thermally with subsequent entropy dilution (see e.g. \cite{Lindner:2010ks,Liao:2010yx}). Typically, the parameter space spanned by the mass (hereafter $m_d$) and active-sterile mixing angle (hereafter $\theta_d$) for sterile neutrino WDM is most tightly constrained by Lyman-$\alpha$ \cite{Boyarsky:2008xj,Lindner:2010ks} and x-ray flux \cite{Boyarsky:2006ag,Boyarsky:2006rp,Boyarsky:2007ay,deVega:2011si,Watson:2011dw} bounds, along with free-streaming, Tremaine-Gunn and big-bang nucleosynthesis bounds, too (see e.g. \cite{Kusenko:2009up, Das:2010ts}). The aggregate effect of these bounds depends on the production mechanism of the sterile neutrino WDM. In particular, at present purely non-resonant production is disfavored, while windows exist for resonant production, production from inflaton decay, or from entropy-diluted thermal freeze out \cite{Boyarsky:2007ay,Boyarsky:2008xj,Boyarsky:2009ix,Kusenko:2009up}.

In this Note, we show that elementary keV \emph{Dirac} sterile neutrinos can be a natural feature of the composite neutrino scenario \cite{ArkaniHamedGrossman:1999,Okui:2004xn,GrossmanTsai:2008,McDonald:2010jm,Duerr:2011ks}, in the same way that the light fermions of the standard model (SM) can arise naturally in the extended technicolor framework \cite{Farhi:1981tc}. Briefly, the composite neutrino scenario is a class of theories in which the right-handed neutrinos are composite bound states of a confining hidden sector (CHS). 

The possibility of such keV sterile neutrinos was first mentioned briefly in Ref. \cite{Grossman:2010iq}, and some of its x-ray flux bounds were investigated in \cite{Hundi:2011et}. In this Note, we present a more generalized discussion of this mechanism that is independent of the precise details of the confining sector, and then proceed to investigate the possible cosmological histories for this WDM candidate. We show certain classes of CHS's can naturally produce keV sterile neutrinos with active-sterile mixing angle in the resonant production window, and a freeze out temperature $\gtrsim$ TeV. Provided the post-inflation reheating temperature is below the TeV scale, then these keV sterile neutrinos could be WDM produced non-thermally via the usual resonant production mechanism \cite{Dodelson:1994sn,Shi:1998km,Fuller:2001sn,Boyarsky:2007ay,Boyarsky:2008xj,Wu:2009yr,Kusenko:2009up, Boyarsky:2009ix,Das:2010ts,Watson:2011dw}, or by a combination of inflaton decay and subsequent non-resonant production \cite{Shaposhnikov:2006xi,Anisimov:2008qs}.

As mentioned above, an alternative to non-thermal WDM production is ultra-relativistic thermal production followed by entropy dilution (see e.g. \cite{Lindner:2010ks}).  This has the advantage of producing colder WDM than resonant production and can better evade the Lyman-$\alpha$ bounds. Usually the diluting entropy is produced by the out-of-equilibrium decay of a sufficiently long-lived heavy particle. In this Note we examine another compelling possibility: The first-order phase transition induced by the confinement of the hidden sector can also produce significant entropy if there is sufficient supercooling. This results in thermal keV WDM. We will discuss the details of this mechanism.

\section{The Composite Dirac Neutrino Model}
\subsection{Setup}
The generic theory of interest is a low-energy effective field theory below a scale $M$. Its group structure is $\Gc\otimes\GF\otimes\GSM$, with $\Gc$ a confining group called $\nu$-color, $\GSM$ the SM gauge groups (or a UV extension), and $\GF$ a global (or weakly gauged) hidden flavor group. The theory consists of three sectors
\begin{equation}
	\chi \sim \Gc\otimes\GF~,\qquad \xi \sim \GF~,\qquad q \sim \GSM\otimes\GF~,
\end{equation}
and which interact only via $M$-scale irrelevant operators.  We call $\chi$ `preons' and say they belong to the CHS. Here $q$ denote the SM fields extended to also carry hidden flavor $\GF$, and we say $\xi$ comprise the `extended hidden sector' (EHS). We assume that the $\chi$ and $\xi$ are purely chiral fermions, but we emphasise that like the SM sector, the $\chi$ and $\xi$ may consist of various different irreps.

The $\nu$-color group confines at a confinement scale $\Lambda \ll M$. Necessarily $M \gg v$, the electroweak scale, so it is convenient to define two parameters
\begin{equation}
	\epsilon \equiv \Lambda/M \ll 1~,\qquad \theta \equiv v/M \ll 1~.
\end{equation}
Confinement of the CHS produces preonic bound states, which we shall crudely denote as $\chi^p$:  The superscript denotes the number of preons participating in the bound state. Formation of a scalar condensate $\chi^{m}$ with $\langle \chi^{m}\rangle \not= 0$ generically induces a spontaneous breaking of the hidden flavor group $\GF \to \GFp\subset \GF$. This produces a new sub-$\Lambda$ effective field theory, which consists of: preonic bound states; $\xi$ and $q$ decomposed into $\GFp$ irreps; and also light `hidden pions'. There are three crucial ideas:

(i) If the CHS has non-trivial $\GFp$ anomalies, then anomaly matching of the CHS to its confined phase, with $\xi$ and $q$ acting as chiral spectators, implies that there are \emph{massless} fermionic bound states after confinement. The remaining bound states generically have masses $\sim \Lambda$, except for the hidden pions, which can be massless or have arbitrarily small masses, depending on the nature of the $\GF$ symmetry breaking. We assume the pion masses are sufficiently small that they make negligible contributions to the DM energy fraction.

Hereafter we shall assume $\GFp = \UF$, and that there are precisely three massless bound states all with the same $\UF$ charge \footnote{In this case decomposition of $q$ under $\GF \to \UF$ could result in multiple copies of SM irreps, also with the same $\UF$ charges, which could be the source of flavor.}. For simplicity we assume the massless bound states have the same number of preons, hereafter denoted $n$, necessarily an odd integer. We shall suggestively denote these bound states as $n_R^i$, $i=1,2,3$ with $\UF$ charge $F(n_R) = +1$. Explicit examples of preonic theories capable of producing such spectra are presented in Ref. \cite{Grossman:2010iq}. The corresponding sub-$\Lambda$ EFT that we shall consider hereafter is shown in Table \ref{tab:SLEFT}. In producing this EFT, we require that the mechanisms of $\GF \to  \UF$ breaking and electroweak symmetry breaking are independent, at least to a good approximation.

\begin{table}[h]
\begin{center}
\begin{tabular*}{0.5\textwidth}{@{\extracolsep{\fill}}c|ccccccc}
	\hline
	\\[-7pt]
	 $\quad$ & $\phi$ & $L^c_L$ & $E_R$ & $Q_L^c$ & $U_R$ & $D_R$ & $n_R$ \\[2pt]
	\hline\hline
	\\[-6pt]
	F & $+1$ & 0 & $-1$ & 0 & $1$ & $-1$ & $+1$\\[2pt]
	\hline
\end{tabular*}
\end{center}
\caption{ $\UF$ charges assignments to the massless bound states $n_R$ and the SM fields $q = \{\phi, Q, U ,D,L,E\}$, which also have the usual SM charges (not shown). The $n_R$ are SM sterile by construction.}
\label{tab:SLEFT}
\end{table}

One can check $2Y-F = B-L$, so $\UF$ is nonanomalous, and the electroweak symmetry breaking (EWSB) pattern is
\begin{equation}
	\label{eqn:SBP}
	\mbox{SU(2)}_{\rm L}\otimes \UY \otimes \UF \to \mbox{U(1)}_{\rm EM}\otimes\mbox{U(1)}_{B-L}~.
\end{equation}
That is, one obtains Dirac neutrinos, with the $n_R$ acting as right-handed neutrinos. Note $\UF$ may be gauged, but we assume its gauge coupling and kinetic mixing with the photon are sufficiently small that they can be neglected. 

(ii) For the sub-$\Lambda$ EFT in Table \ref{tab:SLEFT}, there exist irrelevant operators that couple the preons of the massless $n_R$ -- i.e. the $\Gc$ singlets $\chi^n$ -- to the SM singlet $\bar{L}_L\tilde{\phi}$. Such an operator is generically of form
\begin{equation}
	\label{eqn:SYM}
	\frac{1}{M^{3(n-1)/2}}\bar{L}_L\tilde{\phi}\chi^n \to \epsilon^{3(n-1)/2}\bar{L}_L\tilde{\phi}n_R~,
\end{equation}
after confinement. That is, this operator produces a suppressed Yukawa in the sub-$\Lambda$ EFT. Since $n_R$ are massless and there is $B-L$ symmetry (\ref{eqn:SBP}), this operator leads to light Dirac neutrino masses after EWSB, compared to the electroweak scale.

There may also be other vector-like right-handed fermionic bound states $N_R$ and $N^c_L$, with $F(N_{R,L}) = +1$ We shall again assume for simplicity they contain $n$ preons. Such bound states must form Dirac fermions with $\Lambda$ scale masses, and the $N_R$ will generically also have operators of form (\ref{eqn:SYM}). $N_{R,L}$ are therefore $\Lambda$-scale sterile Dirac neutrinos.

(iii) Under decomposition into $\UF$ irreps, the chiral EHS fields $\xi$ may form real $\UF$ representations and acquire masses. However, because the EHS couples only irrelevantly to the condensate vev $\langle \chi^m \rangle$ responsible for $\GF \to \UF$, the mass terms must be suppressed. This is the same mechanism which suppresses the quark and lepton masses in Extended Technicolor theories \cite{Farhi:1981tc}. Explicitly, for a Dirac fermion $\xi_{R,L}$, such mass terms arise from operators of the form
\begin{equation}
	\label{eqn:EMT}	
	\frac{1}{M^{(3m-2)/2}}\xi \chi^m \xi \to \Lambda\epsilon^{(3m-2)/2}\bar{\xi}_L\xi_R~,
\end{equation}
after confinement \footnote{There may also be mass cross terms involving $\xi_LN_R$, for example. However, we assume that such cross-terms, i.e involving composite and elementary states, are suppressed by the details of the UV theory above $M$. An analogous assumption must also be made for the proton decay operator $uude/M^2$.}. If also $F(\xi_{R,L}) = +1$, then there may exist irrelevant operators that couple the corresponding $\Gc$ singlet $\chi^m\xi$ to $\bar{L}_L\tilde{\phi}$, noting any renormalizable coupling of $\xi$ directly to $\bar{L}_L\tilde{\phi}$ is forbidden by the $\GF$ chiral structure. That is, we could have
\begin{equation}
	\label{eqn:ESM}
	\frac{1}{M^{3m/2}}\bar{L}_L\tilde{\phi}\chi^m\xi \to \epsilon^{3m/2} \bar{L}_L\tilde{\phi}\xi_R~.
\end{equation}
Consequently, such a $\xi_{R,L}$ forms an \emph{elementary} sterile Dirac neutrino with naturally suppressed mass term $\sim \Lambda \epsilon^{(3m-2)/2}$ and coupling to the active sector $\sim \epsilon^{3m/2}$. In principle, there may be several species of such a Dirac neutrino, as well as other EHS fermions with $F \not= \pm 1$ that acquire Dirac or even Majorana masses of the same size.

\subsection{Spectrum} 
We may classify the sub-$\Lambda$ EFT by a tuple $(n,m)$, where $n$ (odd $\ge 3$) is the number of preons in the sterile neutrino bound states, and $m$ (even $\ge 2$) is the number of preons in the symmetry breaking condensate. After EWSB, from eqs. (\ref{eqn:SYM})--(\ref{eqn:ESM}) a $(n,m)$ theory has neutrino mass term,
\begin{equation}
	\Lambda \begin{pmatrix}	\nu_L \\ \xi_L \\ N_L \end{pmatrix}^T \begin{pmatrix} \theta \epsilon^{\frac{3n -5}{2}} & \theta \epsilon^{\frac{3m -2}{2}} & \theta \epsilon^{\frac{3n -5}{2}}\\ 0 &\epsilon^{\frac{3m -2}{2}}  & 0 \\ 0 & 0 & 1\end{pmatrix}\begin{pmatrix}n_R \\ \xi_R \\ N_R\end{pmatrix}~,
\end{equation}
where $\nu_L$ is the SM active neutrino. Each entry of this mass matrix denotes the prefactor of an $\mathcal{O}(1)$ sub-block, whose dimensions depends on the number of species of each type of sterile neutrino. For example, the upper left entry must be $3 \times 3$.

For $m \le n-1$, the mass spectrum can be determined by expansions in $\epsilon$ and $\theta$. One obtains at leading order
\begin{equation}
	\label{eqn:MS}
	m_l \sim v\epsilon^{\frac{3(n-1)}{2}}~,\qquad m_d \sim  \Lambda \epsilon^{\frac{3m-2}{2}}~, \qquad m_h \sim \Lambda~.
\end{equation}
Here the superscripts $l$, $d$ and $h$ denote `light', `dark' and `heavy'.  The left-handed mass basis is, at leading order in $\epsilon$ and $\theta$,
\begin{equation}
	\label{eqn:LHMB}
	\begin{pmatrix} \nu^l_L \\ \nu^d_L \\ \nu^h_L \end{pmatrix}\sim \begin{pmatrix} 1 & \theta & \theta\epsilon^{\frac{3n-5}{2}} \\ \theta  & 1 & \theta^2 \epsilon^{\frac{3n + 6m -9}{2}} \\ \theta\epsilon^{\frac{3n-5}{2}}  & \theta^2\epsilon^{\frac{3n -5}{2}} & 1 \end{pmatrix}\begin{pmatrix} \nu_L \\ \xi_L \\ N_L \end{pmatrix}~,
\end{equation}
and the right-handed mass basis is
\begin{equation}
	\label{eqn:RHMB}
	\begin{pmatrix} \nu^l_R \\ \nu^d_R \\ \nu^h_R \end{pmatrix} \sim \begin{pmatrix} 1 & \theta^2\epsilon^{\frac{3(n-m-1)}{2}} & \theta^2\epsilon^{3n-5} \\ \theta^2\epsilon^{\frac{3(n-m-1)}{2}}  & 1 & \theta^2 \epsilon^{\frac{3n + 3m -7}{2}} \\  \theta^2\epsilon^{3n-5} & \theta^2 \epsilon^{\frac{3n + 3m -7}{2}}& 1 \end{pmatrix}\begin{pmatrix} n_R \\ \xi_R \\ N_R \end{pmatrix}.
\end{equation}
We emphasise that eqs. (\ref{eqn:LHMB}) and (\ref{eqn:RHMB}) denote only sub-block prefactors; the entries of the sub-blocks themselves are generically $\mathcal{O}(1)$ numbers multiplied by the appropriate prefactor.

\begin{figure}[t]
\begin{center}
\includegraphics{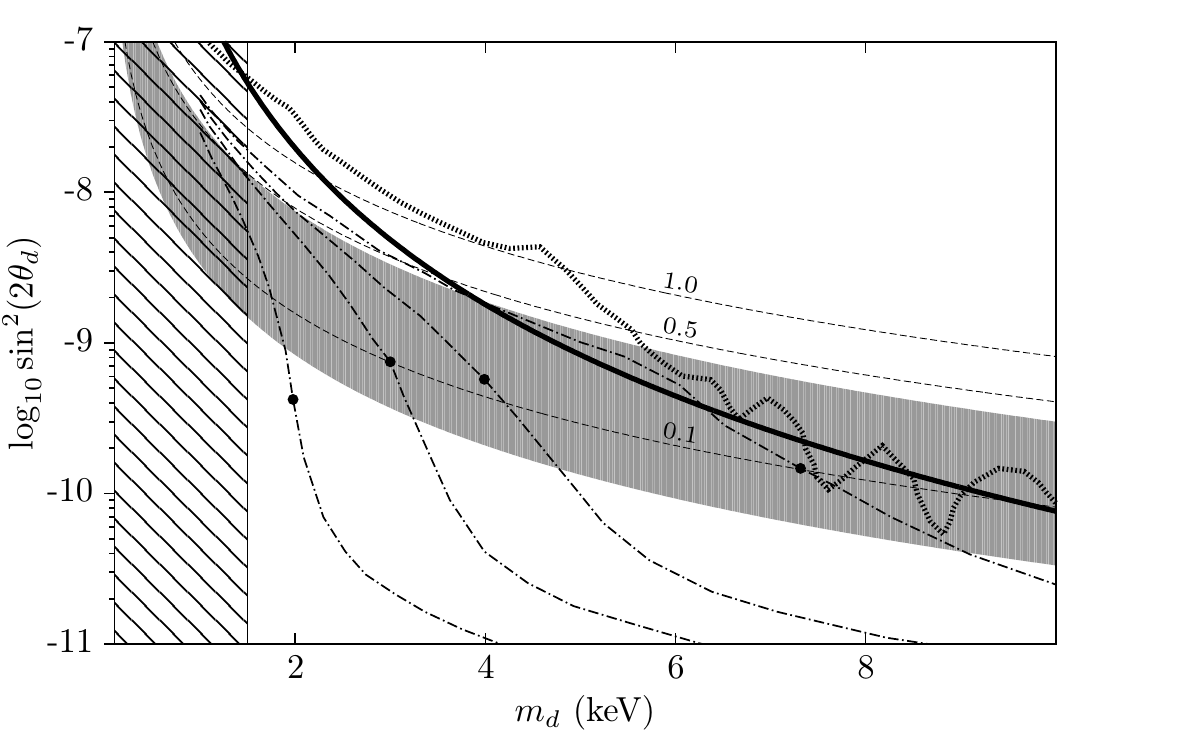}
\end{center}
\caption{Mixing angle $\sin^2(2\theta_d)$ up to $\mathcal{O}(1)$ uncertainty (light gray) as a function of $m_d$, for $(n,n-1)$ theories. Also shown: Non-resonant production contours (dashed lines), labelled by the ratio of $\nu^d$ and DM energy fractions, $\Omega_d/\Omega_{\rm DM}$ \cite{Asaka:2007ls,Boyarsky:2009ix,Kusenko:2009up}; resonant total DM production contours (dash-dotted lines) for lepton asymmetries $Y_{\Delta L} = 8,~12,~16,~25 \times 10^{-6}$ (resp. top to bottom), and their corresponding Lyman-$\alpha$ lower bounds on the WDM mass (black dots) \cite{Boyarsky:2009ix}; the Lyman-$\alpha$ exclusion for thermally produced WDM with subsequent entropy dilution (hatched region, see e.g. \cite{Boyarsky:2008xj,Kusenko:2009up} and eq. (\ref{eqn:LAB}) below) assuming 100\% $\nu^d$ WDM; the x-ray flux exclusion for 100\% $\nu^d$ WDM fitted from most stringent archival data (heavy black line, see e.g. \cite{Boyarsky:2007ay,Boyarsky:2009ix}) and from the most recent observations of dwarf spheriodal galaxies \cite{Loewenstein:2012px} (heavy broken line).}
\label{fig:MTL}
\end{figure}

It is clear from eq. (\ref{eqn:LHMB}) that the dark-active mixing angle $\theta_d \sim \theta$. One can then rearrange eqs. (\ref{eqn:MS}) and (\ref{eqn:LHMB}) into
\begin{equation}
	m_d\theta_d \sim v\bigg(\frac{m_l}{v}\bigg)^{\frac{m}{n-1}}~,\quad \frac{\Lambda}{m_d} \sim \bigg(\frac{m_l}{v}\bigg)^{\frac{2-3m}{3n-3}}~,
\end{equation}
in which the right-hand sides are fully specified by $(n,m)$ and the requirement that $m_l \sim 0.05$ eV, $v \simeq 174$ GeV. Figure \ref{fig:MTL} shows $\sin^2(2\theta_d)$ up to $\mathcal{O}(1)$ uncertainty as a function of $m_d$, with $m = n-1$. Theories with $m < n-1$ have much larger mixing angles, and are therefore ruled out by x-ray flux constraints, so we consider only $(n,n-1)$ theories henceforth. For such theories $M \sim 2\times 10^4(m_d/5~\mbox{keV})$ TeV, and we provide the corresponding $\Lambda$ and $\epsilon$ in Table \ref{tab:CE}.

\begin{table}[h]
\begin{center}
\begin{tabular*}{0.5\textwidth}{@{\extracolsep{\fill}}c|ccc}
	\hline
	\\[-6pt]
	$(n,m)$ & $\Lambda\times (5~ \mbox{keV}/m_d)$ (TeV)&  $\epsilon\times(5~ \mbox{keV}/m_d)$ \\
	\hline\hline
	\\[-6pt]
	$(3,2)$ & $1$  &  $7\times10^{-5}$\\[2pt]
	$(5,4)$ & $10^2$  & $8\times 10^{-3}$\\[2pt]
	$(7,6)$ & $7\times 10^3$ & $9\times 10^{-2}$\\[2pt]
	\hline
\end{tabular*}
\end{center}
\caption{Confinement scale $\Lambda$ and $\epsilon$ for $(n,n-1)$ theories. Such theories with $n>7$ have $\epsilon \not\ll 1$, and are not considered further.}
\label{tab:CE}
\end{table}

It is amusing to note that for the $(n,n-1)$ theories $m_d \sim 5$ keV implies $\sin^2(2\theta_d) \sim 3\times10^{-10}$, which matches the (as yet unconfirmed) \emph{Chandra} results in the Willman I dwarf galaxy \cite{Loewenstein:2009cm}. 

\subsection{Dirac vs Majorana} 
The keV sterile neutrinos in this Note are Dirac, in contrast with the Majorana sterile neutrinos often considered in other WDM scenarios. The WDM production mechanisms that we consider below produce dominantly symmetric DM -- the resonant production mechanism requires an asymmetry in the proper number density $(n_\nu-n_{\bar{\nu}})/n_\nu < 10^{-2}$ \cite{Fuller:2001sn,Wu:2009yr} -- so that the DM particles and antiparticles are present in the same abundances to a very good approximation. The x-ray flux bounds due to sterile neutrinos are therefore insensitive to the mass structure, since decay modes to the active neutrino and antineutrino are present in both cases: I.e, the x-ray flux is due to either $N \to \nu \gamma$ and $N\to \nu^c \gamma$ for a Majorana neutrino $N$, or $\nu^d \to \nu \gamma$ and $\bar{\nu}^d \to \bar{\nu} \gamma$ for the present scenario. Similarly, (non)-resonant production by conversion from the left-handed active neutrinos will produce the same sterile neutrino energy fraction, $\Omega_d$, regardless of the Dirac or Majorana nature of the masses. In Fig. \ref{fig:MTL} we therefore use the existing results for both the x-ray bounds and production processes, without any alteration for the Dirac mass structure.

The x-ray bounds could also be altered by exotic $\nu^d \to X\gamma $ decay channels, that might arise from $M$-scale irrelevant operators. We emphasize that the chiral and composite structure of the composite neutrino framework ensures any such operators are of sufficiently high dimension that the corresponding decay rates are negligible. For example, in the $(3,2)$ theory $\nu^d \to \gamma \nu^l$ or $\nu^d \to \gamma \Pi \nu^l$ could also arise from $\chi^3 \chi^2 F_{\mu\nu}\sigma^{\mu\nu}\xi/M^7$ which confines to $(\Lambda^5/M^7)n_R(\Lambda + \Pi) F_{\mu\nu}\sigma^{\mu\nu}\xi_L$. This respectively produces decay rates $ \sim \epsilon^{12}m_d^3/M^2$ or $ \sim \epsilon^{10}m_d^5/M^4$, that are negligible compared to the decay through mixing with the active neutrinos. 

\subsection{Decoupling} 
Our knowledge of the generic structure of the non-renormalizable operators permits us to consider the cosmological histories of the CHS and EHS, and therefore determine whether the $\nu^d$ sterile neutrinos can be a WDM candidate: satisfying the $(m_d,\theta_d)$ bounds is necessary but not sufficient for this. For the $(n,n-1)$ theories, we now enumerate various important processes and their freeze out temperatures, $T_{\rm fr}$. We assume the effective degrees of freedom at the TeV scale $g_* \sim 10^2$.

(i) $\bar{X}X \leftrightarrow \bar{Y}Y$, where $X,Y \in \{q,\xi,\chi \}$. These processes couple the SM, CHS and EHS. The dimension-5 operator $\phi^\dagger \phi \bar{X}X$ is heavily suppressed, since $X$ are all chiral. The leading operators are then the dimension-6
\begin{equation}
	\frac{1}{M^2}\bar{X}\gamma^\mu X \bar{Y} \gamma_\mu Y~;~T_{\rm fr} \sim \bigg[\frac{g_*^{\frac{1}{2}}M^4}{\Mpl}\bigg]^{1/3} \sim~\mbox{TeV}~,
\end{equation}
and similarly for $\phi^\dagger \partial_\mu \phi\bar{X}\gamma^\mu X/M^2$.  Note that the current collider constraint on the dark matter - quark interaction is insensitive to the coupling due to the large mediator mass, $M$ \cite{Rajaraman:2011uq,Fox:2011fk}.

(ii) $\bar{\xi}_R\xi_L \leftrightarrow 2\Pi$, where $\Pi$ denotes the hidden pions. This process is generated by the non-linear sigma operator
\begin{equation}
	\label{eqn:XXPP}
	m_d\bar{\xi}_R\xi_L e^{i\Pi/\Lambda}~;~T_{\rm fr} \sim \bigg[\frac{g_*^{\frac{1}{2}}\Lambda^4}{(m_d)^2\Mpl}\bigg] \sim~\mbox{TeV}~,
\end{equation}
for the $(3,2)$ theory, and much larger for $(5,4)$ and $(7,6)$.

(iii) $\bar{\nu}^d_L\nu_L^d \leftrightarrow \bar{q}q$. This can occur also through $W$ and $Z$ exchange, and must freeze out before BBN. The pertinent operators are
\begin{equation*}
	\frac{g(\theta_{d})^2}{2c_W}\bar{\nu}^d_L\slashed{Z}\nu^d_L~,~\frac{g\theta_d}{\sqrt{2}}\bar{\nu}^d_L\slashed{W}\ell_L~;~T_{\rm fr} \!\sim \!\bigg[\frac{g_*^{\frac{1}{2}}m_W^4}{(\theta_{d})^4\Mpl}\bigg]^{1/3} \!\!\!\!\!\! \sim~\mbox{TeV}~.
\end{equation*}

(iv) $\bar{\nu}^l_L\nu^{l}_R \leftrightarrow 2\Pi$. This must also freeze out before the BBN epoch. The non-linear sigma coupling of $\nu^l_{L,R}$ to the hidden pions is suppressed by both the left and right mixing between active and sterile sectors. From eqs. (\ref{eqn:LHMB}) and (\ref{eqn:RHMB}) this leads to an extra prefactor of $(\theta_d)^3$ for the non-linear sigma operator in eq. (\ref{eqn:XXPP}), and therefore a decoupling much larger than the TeV scale.

(v) $2h \leftrightarrow 2\Pi$. This is generated by the operator $\phi^\dagger\phi (\chi^m)^\dagger \chi^m/M^{3m-2}$ which confines to the dimension-4 operator $\epsilon^{3m-2}\phi^\dagger\phi \Pi\Pi$. This becomes efficient only \emph{below}
\begin{equation}
	T_{\rm fr} \sim \epsilon^{6m-4}\Mpl/ g_*^{\frac{1}{2}} \lesssim 10^{-7}~\mbox{eV}~,
\end{equation}
for $(n,n-1)$ theories, and therefore does not produce significant recoupling.

\section{Warm Dark Matter}

\subsection{Non-Thermal WDM} 
The moral of the above analysis is that approximately below the TeV scale, the SM, CHS and EHS are decoupled. From Table. \ref{tab:CE}, confinement of the CHS also occurs at latest at the TeV scale. As a result, we may imagine a scenario in which the post-inflation reheating temperature $T_{\rm rh} <$ TeV. In this case, the sterile Dirac neutrinos $\nu^d$ might never be in thermal contact with the SM plasma, and therefore be produced non-thermally through the (non)-resonant production mechanism \cite{Dodelson:1994sn,Shi:1998km,Fuller:2001sn,Wu:2009yr}, forming the WDM.

As can be seen in Fig. \ref{fig:MTL}, the predicted $(m_d,\theta_d)$ values fall outside the $\Omega_d > \Omega_{\textrm{DM}}/2$ non-resonant production region, which itself is ruled out by the combination of Lyman-$\alpha$ \cite{Seljak:2006qw,Boyarsky:2008xj} and x-ray flux bounds \cite{Boyarsky:2007ay, Boyarsky:2009ix}. However the $(m_d,\theta_d)$ ranges still overlap an allowed window for full WDM resonant production if there is a sufficiently large lepton asymmetry \cite{Fuller:2001sn,Laine:2008pg,Wu:2009yr,Kusenko:2009up,Boyarsky:2009ix}. Alternatively, in this low reheat scenario, coupling of the sterile neutrinos to the inflaton -- an SM singlet -- could result in significant non-thermal WDM production from its decay \cite{Shaposhnikov:2006xi,Anisimov:2008qs}, with the remaining fraction (if any) produced by non-resonant production.

Just as for exotic x-ray decay channels, the chiral structure of the SM, CHS and EHS generically suppresses the operators that may produce non-thermal sterile neutrino WDM from SM decays. For example, from eqs. (\ref{eqn:ESM}) and (\ref{eqn:MS}) it is clear that Higgs to sterile neutrino decay rate is suppressed by $\epsilon^{3m} \sim (m_l/v)^2$ for $(n,n-1)$ theories, so there is no significant production from the Higgs decay channel. Along similar lines to the inflaton scenario, one might putatively extend the Higgs sector with a SM singlet that can decay to the EHS  without such suppression (see e.g. \cite{Kusenko:2006rh,Petraki:2007gq}), however we do not make any such assumptions about the SM Higgs sector here.

One might also consider production via lepton or hadron decays such as $\tau \to e\xi\xi$ or $B \to K\xi\xi$ respectively. The chiral structure ensures such processes can only be mediated by operators of the form
\begin{equation}
	\frac{\lambda_{ij}}{M^2}\bar{q}^i\gamma^\mu q^j\bar{\xi}\gamma_\mu\xi~.
\end{equation}
This type of operator necessarily produces FCNCs, too, but the large mediator scale $M$ easily evades the present bounds for quark FCNCs \cite{pdg:2012}. One finds for the dominant top decay process $\Gamma/H(m_t) \lesssim 10^{-4}$. For semi-relativistic tops in thermal equilibrium, this produces a sterile neutrino energy fraction $\Omega_d \sim 1\% ~ \Omega_{\textrm{DM}}$, so that this production channel can be neglected.  Similarly, production from spin-1 bound state decays like $\rho_0 \to \xi\xi$ is negligible due to suppression of the rate by a $(\Lambda_{\textrm{qcd}}/M)^4$ factor.

\subsection{Thermal WDM} 
The $(3,2)$ theory exhibits the interesting feature that the decoupling temperature of the EHS, $T_d$, the confinement temperature of the CHS, $T_c \sim \Lambda$, and decoupling of temperature the CHS, $T_\chi$, all occur at the TeV scale.  In contrast to the non-thermal resonant scenario, for a $(3,2)$ theory one may plausibly consider a scenario in which all three sectors are initially in thermodynamic equilibrium, the lepton asymmetry is small, and 
\begin{equation}
	\label{eqn:OS}
	T_d > T_c > T_\chi~.
\end{equation}

In this scenario, the EHS fermions $\xi$ freeze-out ultra-relativisitically before confinement, and there is no subsequent resonant production: from Fig. \ref{fig:MTL} we see that fractional non-resonant production at the $10\%~\Omega_{\textrm{DM}}$ level may still occur, but we shall neglect this henceforth as it is a subdominant contribution. Defining $Y \equiv n/s$ -- the ratio of the comoving number density and entropy density -- then for each \emph{Dirac} $\xi$ species 
\begin{equation}
	Y_{\xi} =  \frac{135\zeta(3)}{2\pi^4}\frac{1}{\gs^d}~,
\end{equation}
where $\gs^d$ is entropic effective equilibrium number of degrees of freedom at freeze-out. 

Even if only one species of $\xi$ -- the Dirac $\xi_{R,L}$ -- obtains a mass $m_d$, which we assume henceforth, such a $Y_\xi$ leads to over-closure unless $\gs^d \sim 10^4$. This is unnaturally large since $\gs \sim 10^2$ for the SM at this scale. However, if after freeze-out the entropy increases by a factor $\gamma$, then the frozen out species are diluted, $Y_{\xi} \to Y_\xi/\gamma$.  The present-day energy fraction for the Dirac $\nu^d$, which are an admixture dominantly composed of $\xi_{R,L}$, is then
\begin{equation}
	\label{eqn:PDEF}
	\frac{\Omega_{d}}{\Omega_{\rm DM}} \simeq \frac{Y_\xi m_d s_0}{\rho_c\Omega_{\rm DM}} = \frac{1.1\times 10^4}{\gs^d\gamma}\bigg(\frac{m_d}{5~\mbox{keV}}\bigg)~,
\end{equation}
in which we used $s_0 \simeq 2.89\times10^3$ cm$^{-3}$, $\rho_c \simeq 10.5 h^{2}$cm$^{-3}$keV, and $\Omega_{\rm DM} = 0.105h^{-2}$. It is clear that we need $\gs^d\gamma \gtrsim 10^4$ for a DM candidate. 

\subsection{Supercooled Confinement} 
The ordering (\ref{eqn:OS}) permits us to consider the confinement of the CHS as the source of entropy that dilutes $Y_\xi$ after freeze-out. The entropy production from a confinement-induced first-order phase transition can be significant if it occurs suddenly after supercooling \cite{DeGrand:1984sc,Csorgo:1994dd}. That is, if the confinement phase transition (CPT) begins at a cooler temperature $T_i < T_c$, and the duration of the transition $\tau_c \ll 1/H(T_i)$, the Hubble time at temperature $T_i$. 

Before confinement -- at temperature $T_i$ -- and after confinement -- at temperature $T_f>T_\chi$ --, we suppose that we have equilibrium plasmas. By construction
\begin{align}
	\gs(T_i) & \equiv \gs^i = \gs^{\rm SM} + \gs^{\rm c}  \simeq 2\times10^2~,\notag\\
	 \gs(T_f) & \equiv  \gs^f  \equiv \gs^{\rm SM} + \gs^{\rm bs} \simeq 10^2~.
\end{align}
Here  $\gs^{\rm SM}$, $\gs^{\rm c}$ and $\gs^{\rm bs}$ denote the effective equilibrium relativistic degrees of freedom in the SM, CHS and the bound states. By construction, for three $n_R$ we have $\gs^{\rm bs}=2\cdot3\cdot(7/8)+N_{\Pi}$ with $N_{\Pi}$ the number of hidden pions. We have assumed $\gs^{\rm bs} \sim 10$ and $\gs^{\rm SM},~\gs^{\rm c} \simeq 10^2$. Note that since the frozen out $\xi_{L,R}$ have only four degrees of freedom, then $\gs^d \simeq \gs^i$.

Since $T_f>T_\chi$, then such entropy production leads to reheating of \emph{both} the CHS and SM, because they only decouple later at $T_\chi$. This mutual reheating means the present DM temperature, $T^0_{d}$, compared to that of the active neutrinos, $T^0_\nu$, is just 
\begin{equation}
	\label{eqn:TRDM}
	\frac{T^0_{d}}{T^0_\nu} = \bigg(\frac{\gs^f}{\gamma\gs^d}\frac{\gs^\nu}{\gs^{\rm SM}}\bigg)^{1/3}\simeq \bigg(\frac{10.75}{1.1\times10^4(m_d/5~\mbox{keV})}\bigg)^{1/3}~,
\end{equation}
from eq. (\ref{eqn:PDEF}) and since $\gs^{f} \simeq \gs^{\rm SM}$. Equation (\ref{eqn:TRDM}) implies the entropy-diluted thermal WDM is red-shifted compared to the active neutrino plasma. The Lyman-$\alpha$ bounds \cite{Seljak:2006qw,Boyarsky:2008xj,Lindner:2010ks} require non-resonantly produced WDM -- at present temperature $T^0_\nu$ -- to satisfy $m_\textrm{nrp} > 10$ keV. Since the free-streaming length $\lambda_{\textrm{FS}} \propto T/m$ (see e.g. \cite{Kusenko:2009up}), this Lyman-$\alpha$ bound translates to $m_d  > 10(T^0_d/T^0_\nu)$ keV. Together with eq. (\ref{eqn:TRDM}) we find that thermally produced $\nu^d$ may safely avoid the Lyman-$\alpha$ bound, provided 
\begin{equation}
	\label{eqn:LAB}
	m_d>1.5~ \mbox{keV}~. 
\end{equation}
This is the Lyman-$\alpha$ bound displayed in Fig. \ref{fig:MTL}. 

Note also that the $n_R$ and hidden pion contribution to the effective number of neutrino degrees of freedom, $\delta N^{\rm eff}_\nu$, at the big-bang nucleosynthesis (BBN) epoch is
\begin{equation}
	\delta N^{\rm eff}_\nu = (8/14)\gs^{\rm bs}\big(\gs^\nu/\gs^{\rm SM}\big)^{4/3} \lesssim 0.26(\gs^{\rm bs}/10)~.
\end{equation}
It is amusing to note that the right-handed neutrinos together with the hidden pions can supply sufficient effective degrees of freedom at the BBN epoch to significantly contribute to the observed $\delta N^{\rm eff}_\nu \sim 1$ excess (see e.g \cite{Komatsu:2010fb,Benson:2011ut}). In contrast, this is difficult to achieve with seesaw models, or even ad hoc Dirac neutrino models.

\subsection{Entropy Production Estimate} 
The massive bound states typically have masses $x\Lambda$, with $x\gtrsim 1$, so they are non-relativistic. Their corresponding widths are generically also $\Gamma \sim\Lambda$. This leads to $\Gamma/H(T_i) \sim \Mpl\Lambda /T_i^2 \ggg 1$. In contrast, the longest-lived heavy bound state we could contemplate decays only via exchange of an $M$-scale boson, like the electroweak decay of the $\Lambda^0$ baryon of QCD. In this case, the decay rate is $\Gamma \sim \Lambda x^5 \epsilon^4$. For the $(3,2)$ theory $\epsilon \sim 10^{-4}$, so that $\Gamma/H(T_i) \gtrsim x^5 \epsilon^4 \Mpl\Lambda/T_i^2 \gg 1$. This means that even for a sudden CPT, the heavy bounds states all decay within $\tau_c$ and generically, predominantly produce hidden pions and $n_R$ with energies $\sim T_c$. It seems reasonable, then, to treat the CPT as a quasiequilibrium process, in which the non-relativistic heavy bound states have exponentially suppressed number and energy densities, while pions and $n_R$ are thermal with temperature $T_c$.

\begin{figure}[t]
\begin{center}
\includegraphics{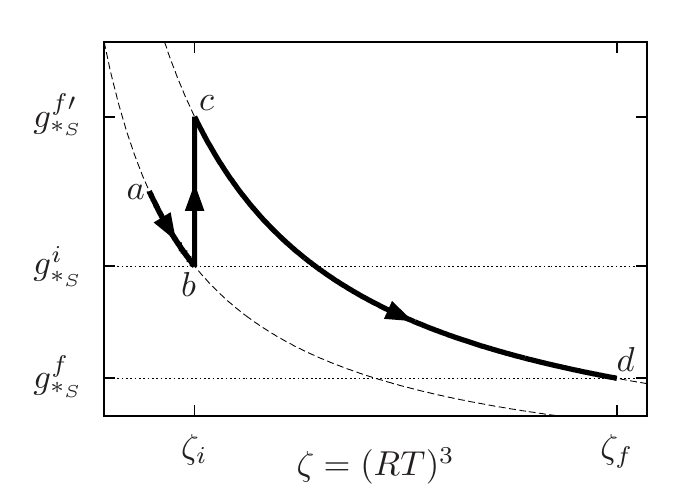}
\end{center}
\caption{A sketch of the thermal history. Species freeze-out ($a$-$b$) along the $S_i$ adiabat (lower dashed), is followed by the CPT ($b$-$c$), which is a first-order $\gs$ phase transition in $\zeta$. The CPT is followed by thermalization (c-d) along the $S_f$ adiabat (upper dashed) until $\gs = \gs^f$ at which $T= T_f$. Once $T = T_\chi$, the CHS and SM decouple.}
\label{fig:CH}
\end{figure}

With this in mind, one can estimate the amount of entropy production by treating the CPT as a first-order phase transition in $\gs$, as a function of $\zeta \equiv (RT)^3$. Here $R$ is the universe scale factor and $T$ the equilibrium temperature.  The picture is that confinement begins at supercooled plasma temperature $T_i$, and suddenly produces the relativistic pions and $n_R$ at temperature $T_c$, so that $\gs$ undergoes a jump at $\zeta_i = (R_iT_i)^3$ from $\gs^i$ to 
\begin{equation}
	\label{eqn:GSF}
	\gs^{f\prime} = \gs^{\rm SM} + \gs^{\rm \rm bs}\big(T_c/T_i\big)^3~.
\end{equation}
This expression for $\gs^{f\prime}$ follows just from the definition $\gs(T) \equiv \sum_\alpha \gs^\alpha(T_\alpha/T)^3$, a sum over species at different temperatures. After the phase transition, the plasma undergoes an adiabatic thermalization until $\gs = \gs^f$ and $T = T_f$. SM-CHS decoupling at $T_\chi$ follows thereafter. Figure \ref{fig:CH} shows this history. 

Provided $(T_c/T_i)^3 \gg \gs^{\rm SM}/\gs^{\rm bs} \sim 10$, the entropy production estimate from eq. (\ref{eqn:GSF}) is then 
\begin{equation}
	\gamma \equiv \frac{S_f}{S_i} = \frac{\gs^{f\prime}\zeta}{\gs^i\zeta} \simeq \frac{\gs^{\rm \rm bs}}{\gs^i}\bigg(\frac{T_c}{T_i}\bigg)^3~.
\end{equation}
The important feature of this na\"\i ve estimate is the $(T_c/T_i)^3$ dependence of the entropy production. A more careful treatment in Ref. \cite{DeGrand:1984sc} produces the result
\begin{equation}
	\gamma \simeq \frac{1}{r}\bigg(\frac{r-1}{3}\bigg)^{3/4} \bigg(\frac{T_c}{T_i}\bigg)^3~,\qquad r \equiv \frac{\gs^i}{\gs^f}~.
\end{equation}
One also finds $T_f = [(r-1)/3]^{1/4}T_c$. Using this result and eq. (\ref{eqn:PDEF}), and fixing $r=2$, it follows that for $\Omega_d \le \Omega_{\rm DM}$ (i.e. $\gamma \gs^d \ge1.1\times10^4 m^d/5~\mbox{keV}$) we require
\begin{equation}
	\label{eqn:TCTI}
	\frac{T_c}{T_i} \ge 6.3 \bigg(\frac{2\times10^2}{\gs^{d}}\bigg)^{1/3}\bigg(\frac{m_d}{5~\mbox{keV}}\bigg)^{1/3}~.
\end{equation}
Note $T_f = 0.76T_c$ here, so it is plausible that $T_f > T_\chi$. By comparison to eq. (\ref{eqn:TCTI}), the QCD maximal supercooling is $T_c/T_i \simeq 1.7$ \cite{DeGrand:1984sc}. However, given that this upper bound will be sensistive e.g. to the tunneling probabilities between the metastable ($\GF$ symmetric) and stable ($\GFp$ symmetric) vacua, the degree of supercooling required in this estimate is not implausible. 

\section{Conclusions} 
Within the composite neutrino framework, we have shown in this Note that keV sterile Dirac neutrinos can be naturally produced with mixing angles appropriate for non-thermal resonant production, provided the composite neutrinos are all comprised of $n$ preons and the scalar condensate vev has $n-1$ of them. Alternatively, for a $(3,2)$ theory, a single keV sterile Dirac neutrino species could be WDM produced by entropy-diluted ultrarelativistic freeze-out. In this latter case the entropy can be provided by a supercooled confinement-induced phase transition.

\acknowledgments
The authors thank Kfir Blum, Yuval Grossman, Roni Harnik, Bibhushan Shakya and Tomer Volansky for helpful discussions. This work is supported by the U.S. National Science Foundation through grant PHY-0757868.


\end{document}